# Optimization of radiation pressure in dielectric nanowaveguides

**Janderson Rocha Rodrigues[1,2,*] and Vilson Rosa Almeida[1,2,3]**

[1]*Instituto Tecnológico de Aeronáutica, 50 Praça Marechal Eduardo Gomes, Vila das Acacias, São José dos Campos - SP. 12228-900, Brazil*
[2]*Instituto de Estudos Avançados, 01 Trevo Cel Av José Alberto Albano do Amarante, Putim, São José dos Campos - SP. 12228-001, Brazil*
[3]*Instituto Científico e Tecnológico da Universidade Brasil, 235 Rua Carolina Fonseca, São Paulo, SP, 08230-000, Brazil*
[*]*jrr@ita.br*

**Stimulated Brillouin scattering (SBS) processes have been allowing important technological breakthroughs in integrated photonics and nano-optomechanics, by exploiting light-sound (photon-phonon) interactions at the nanoscale. These nonlinear processes are created by two main effects: radiation pressure and electrostriction; however, the former is the predominant one in high-index-contrast nanowaveguides. In this letter, we derive a simple set of analytical expressions that can be used for optimizing the radiation pressure on the waveguide boundaries, for any optical mode, polarization, and wavelength. We observe a very strong influence of the waveguide geometric parameters on the optimal radiation pressure value. Furthermore, we explain how the existence of such optimal geometric dimensions is physically related to the minimization of the electromagnetic momentum flow in the propagation direction. This work provides a novel and robust method, yet simple, to optimize the radiation pressure in dielectric nanowaveguides, which may be of great relevance for designing integrated photonic-phononic devices.**

Light and sound multiphysics interactions can be efficiently explored in dielectric devices by using stimulated Brillouin scattering (SBS) process[1-3]. In the last decades, the suppression and enhancement of Brillouin scattering have been extensively studied in optical fibers[1]. In conventional optical fibers, SBS effect is mainly generated by electrostrictive forces induced by light – electrostriction is a bulk effect that occurs in dielectric materials, related to the deformation of their original geometric shapes due to the application of an electric field[1]. Electrostriction effect is energetically linked to its reciprocal process called the photoelastic effect, which describes changes in the material optical properties due to a mechanical deformation[4-9]. However, besides of the electrostrictive force, another kind of optical force becomes relevant to the SBS effect when the waveguides' dimensions are reduced to the nanoscale – the radiation-pressure induced force[4-10]. Radiation pressure exerts forces on the waveguide boundaries due to the intrinsic physical nature of an optical mode, by means of multiple reflections (momentum exchange) on the internal dielectric interfaces[9]. This surface effect is also energetically linked to its reciprocal process called moving-boundary effect, which describes the change in the waveguide cross section due to the action of a transversal force[4-11]. In particular, Rakich *et* al. have shown that, in nanowaveguides, radiation pressure can achieve very high values and, therefore, become the dominant optical force in SBS effect, overcoming electrostriction effect[4,5,10]. Waveguides' nanoscale dimensions and high-index contrast are responsible for the high confinement of light and sound, resulting in a dramatic enhancement on the SBS gain, much far beyond of what is obtained solely through their intrinsic materials nonlinearities, opening new opportunities for acousto-optics interactions in nanophotonics[5,12]. Such theoretical predictions have been closely followed by experimental demonstrations of SBS effects on different nanophotonic and nano-optomechanical waveguides[13-20]. Furthermore, novel SBS-based on-chip signal-processing technologies have been introduced, including recent experimental demonstrations of receiver/emitters [21], memories[22], and lasers [23-25], among many others[26].

Here, we developed a new, simple, and general method to obtain the waveguide's dimensional parameters (e.g., height and/or width) that maximizes the radiation pressure on its surface, for any order mode, polarization, and wavelength. We derive analytical expressions capable of computing this specific dimension using only one line of code, saving a tremendous amount of computational effort and time, commonly demanded during the optimization design, by employing either FDTD (*Finite-Difference Time-Domain*) or FEM (*Finite-Element Method*) numerical tools. We noticed that, for a given combination of materials' refractive indexes (cladding and core) and wavelength, there is solely one specific waveguide set of dimensional parameters that corresponds to a unique (global) maximum point of radiation pressure, which changes accordingly with the optical mode order and polarization. As examples of our findings, we apply our method to two very distinct waveguide geometries: a rib silicon waveguide suspended in air and a strip chalcogenide waveguide buried in silicon dioxide, and exact results are obtained for both cases. Finally, we prove that the existence of a maximum point of radiation pressure on the waveguide surfaces is generated by the minimization of longitudinal momentum flow, which besides of explaining the accuracy of our method, also physically explain the origin of such optimal point – this approach, to the best of our knowledge, has not been presented before.

Consider the dielectric structures schematically shown in Fig. 1 (a)-(c). These structures are formed by a high-index material waveguide ($n_H$) and surrounded by a low-index material medium ($n_L$). Figure 1(a) shows a 3D view of the rectangular cross-section waveguide, with width $w$ and height $h$, as illustrated in the panel (b), which is invariant along the propagation direction ($z$-direction) and delimited by the length $L$.

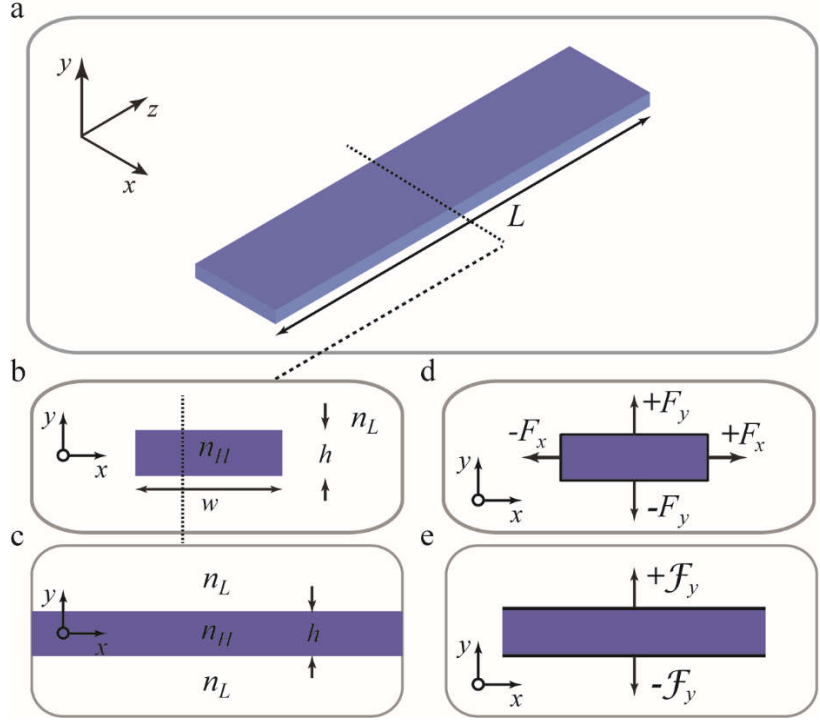

**Fig. 1. Schematic views of the dielectric waveguides and diagram of forces.** (a) 3D view of the dielectric waveguide, with a rectangular cross-section and composed by a high-index material ($n_H$) core surrounded by a low-index material ($n_L$). (b) cross-section (2D) view of the rectangular waveguide with width $w$ and height $h$. (c) planar (1D) approximation by considering the structure's invariance also in the $x$-direction. (d) and (e) show the schematics of radiation-pressure induced optical forces on the surfaces of the rectangular ($F_x, F_y$) and the planar ($\mathcal{F}_y$) waveguides.

Our analytical models are developed, without loss of generality, by considering the symmetric planar (slab) waveguide shown in Fig. 1(c), which is translational invariant in both the transversal $x$-direction and in the propagation $z$-direction). We assume that the planar waveguide is excited by a harmonic source with a fixed wavelength $\lambda_0$ and a vacuum wavevector of magnitude $k_0$, where $k_0 = 2\pi/\lambda_0$. Maxwell's equations, under these assumptions, lead to two independent sets of solutions for the field polarization: Transverse Electric (TE), where the electric field is polarized along the $x$-axis ($E_x, H_y, H_z$) or Transverse Magnetic (TM), where the magnetic field is polarized along the $x$-axis ($E_y, E_z, H_x$). Due to the structure symmetry, with respect to the $x$-axis, the electromagnetic field distributions for each mode can be classified into symmetric/even or antisymmetric/odd. The well-known transcendental (characteristic) equation for the TE symmetric eigenmodes is given by

$$\tan\left(\kappa_H \frac{h}{2}\right) = \frac{\gamma_L}{\kappa_H} \tag{1}$$

where $\kappa_H$ and $\gamma_L$ are the transversal wavevector and the field decay coefficient, respectively, given by $\kappa_H = k_0\sqrt{n_H{}^2 - n_{eff}{}^2}$ and $\gamma_L = k_0\sqrt{n_{eff}{}^2 - n_L{}^2}$, where $n_{eff}$ is the mode's effective index. The TM symmetric transcendental equation is obtained by multiplying the right-hand side of Eq. (1) by the ratio $n_H^2/n_L^2$. Furthermore, the antisymmetric versions of these equations are obtained by inverting their respective right-hand sides and multiplying them by $-1$ (presented in Supplementay Information - Section A).

On the other hand, the radiation pressure $\boldsymbol{p}$ on waveguide's boundaries $\partial wg$ for a given mode is given by[10]:

$$\oint_{\partial wg} \boldsymbol{p} \cdot \boldsymbol{r} \mathrm{d}l = \frac{P}{c}(n_g - n_{eff}) \tag{2}$$

where $n_g$ is the group index, $P$ is the optical power, and $c$ is the speed of light in vacuum. As a result, in the case of a dielectric waveguide with a rectangular cross-section, as shown in Fig. 1(b), the radiation pressure takes the form:

$$\bar{p}_x + \bar{p}_y = \frac{P}{c} \frac{(n_g - n_{eff})}{A_{wg}} \tag{3}$$

where $\bar{p}_x$ and $\bar{p}_y$ are the spatial-averaged radiation pressure in the $x$- and $y$-direction, given by $\bar{p}_x = p_x/h = F_x/Lh$ and $\bar{p}_y = p_y/w = F_y/Lw$, respectively, $A_{wg} = wh$ is the waveguide cross-section area, $F_x$ ($F_y$) is the radiation-pressure induced optical force in the $x$-direction ($y$-direction), as schematically show in Fig. 1(d), and the total optical force is given by $F_x + F_y$. Moreover, by applying the RTOF (*Response theory of optical forces*) method, it is possible to express each component of the optical force by the derivative of the effective index as[27-28],

$$F_q = \frac{PL}{c} \frac{dn_{eff}}{dq} \tag{4}$$

where $q$ is a generalized coordinate. The $x$-component ($y$-component) of the force $F_x$ ($F_y$) is computed by doing $dq = dw$ ($dq = dh$). Both methods represented in Eqs. (2) and (4) have been shown to be numerically equivalent to the Maxwell Stress Tensor (MST) formalism[27-28]. Particularly, in the case of the planar waveguide represented in Fig. 1(c), as its width tends to infinity ($w \to \infty$), the $x$-component of the optical force goes to zero ($F_x = 0$), and the optical force per unit length in the $x$-direction is given by

$$\mathcal{F}_y = \frac{\mathcal{P}L}{c} \frac{dn_{eff}}{dh} \tag{5}$$

where the ratio $\mathcal{P}$ is the mode optical power per unit length in the $x$-direction. In a limiting case, where the rectangular waveguide can be approximated by a planar one ($w \gg h$), the optical forces and powers on the two waveguides are related by $F_y = \mathcal{F}_y w$ and $P = \mathcal{P}w$.

From a mathematical point of view, due its transcendental nature presented in Eq. (1), the modes' effective indexes have no closed-form solution. However, by solving it numerically (or graphically), it is possible to notice that it has a sigmoid function behavior, which monotonically increases from one asymptote, bounded below by $n_L$, to the other, limited to $n_H$, as height goes from zero to infinity; this behavior is shown in Fig. 2(a) and (b), for the fundamental TE and TM modes ($TE_0$ and $TM_0$) on a silicon planar waveguide ($n_H = 3.4764$) in air ($n_L = 1.0003$), at a wavelength of $\lambda_0 = 1550$ nm. This kind of curve has a unique non-stationary point of inflection, that occurs at a specific value of height, and its first derivative has a bell shape format with a maximum point that corresponds to the maximum value of optical force, accordingly to Eq. (5). By performing the derivative on the transcendental equation (Eq. (1)) with respect to the planar waveguide's height, $h$, we obtain

$$\frac{dn_{eff}}{dh} = \frac{n_H^2 - n_{eff}^2}{n_{eff}h + 2n_{eff}\big/\left(k_0\sqrt{n_{eff}^2 - n_L^2}\right)} \tag{6}$$

Substituting Eq. (6) into Eq. (5), and considering the universal condition for guides modes in channel and slab waveguides ($n_L < n_{eff} < n_H$), ensues that the optical force induced by radiation pressure is always positive and tends to push the waveguide's interface apart, i.e., it is an expansion force, which is in agreement with previous results [4]. Additionally, the waveguide's height which maximizes the radiation pressure on its surfaces may be obtained by deriving Eq. (6), with respect to $h$, and equaling it to zero, i.e, $d\mathcal{F}_y/dh = 0$; solving for the optimal height, we obtain,

$$h^{TE} = \frac{\left[\frac{2}{\gamma_L^2} - \frac{2}{k_0^2 n_{eff}^2} - \frac{6}{\kappa_H^2}\right]}{\left[\frac{1}{k_0^2 n_{eff}^2} + \frac{3}{\kappa_H^2}\right]} \frac{1}{\gamma_L} \tag{7}$$

Replacing $h^{TE}$ into Eq. (6), and then the result into Eq. (4), we obtain the maximum value of the optical force, $\mathcal{F}_y$, for the symmetric TE modes, for a given power $\mathcal{P}$ and $\lambda_0$. Similarly, by repeating the same procedure for the TM modes, we obtain

$$h^{TM} = \frac{\left[\frac{4X}{(Xn_{eff}^2 - 1)k_0^2} + \frac{2}{\gamma_L^2} - \frac{2}{k_0^2 n_{eff}^2} - \frac{6}{\kappa_H^2}\right]}{\left[\frac{1}{k_0^2 n_{eff}^2} + \frac{3}{\kappa_H^2}\right](Xn_{eff}^2 - 1)} \frac{1}{\gamma_L} \tag{8}$$

where $X$ is a refractive-indexes related constant, given by $X = 1/n_L^2 + 1/n_H^2$. Since, the effective index is an implicit function of the height, the results presented in the Eqs. (7) and (8) still preserve their transcendental characteristics. Nevertheless, by substituting Eq. (7) into (1), and easily solving it numerically, we obtain the planar waveguide's height that maximizes the optical force for each symmetric mode with TE polarization. Additionally, the height that maximize the radiation pressures for each antisymmetric mode may be obtained by substituting exactly Eq. (7) on the antisymmetric version of Eq. (1); similar procedures should be followed for the TM polarization counterparts.

On one hand, Figs. 2(a) and (b) show the optical forces per unit length ($\mathcal{F}_y$) applied over the horizontal interfaces of a silicon planar waveguide, as a function of its height, as well as their highlighted optimal heights (where the maximum forces occur), for both the fundamental TE and TM modes at 1550 nm for $\mathcal{P} = 1$ mW/μm . The optimized height for the TE polarization, $h^{TE} = 52.04$ nm, is lower than the TM one, $h^{TM} = 225.8$ nm, which is explained mainly due to the positive extra term in Eq. (8) numerator, when compared with Eq. (7), as well as to the fact that, for a given height, the effective indexes of the TM modes are lower than those for their TE counterparts. Furthermore, at these optimized force values, the TM polarization generates a stronger maximum optical force ($\mathcal{F}_y(h^{TE}) = 45.7$ pN/μm²/mW) than the TE one ($\mathcal{F}_y(h^{TM}) = 44.2$ pN/μm²/mW), due to the large discontinuity of the electric field component ($E_y$) at the planar interface in the former polarization. However, it is interesting to notice that the TE polarization still generates a reasonable strong force, even though it has neither discontinuity in the electric field component ($E_x$), nor it is orientated on the same direction of the optical force.

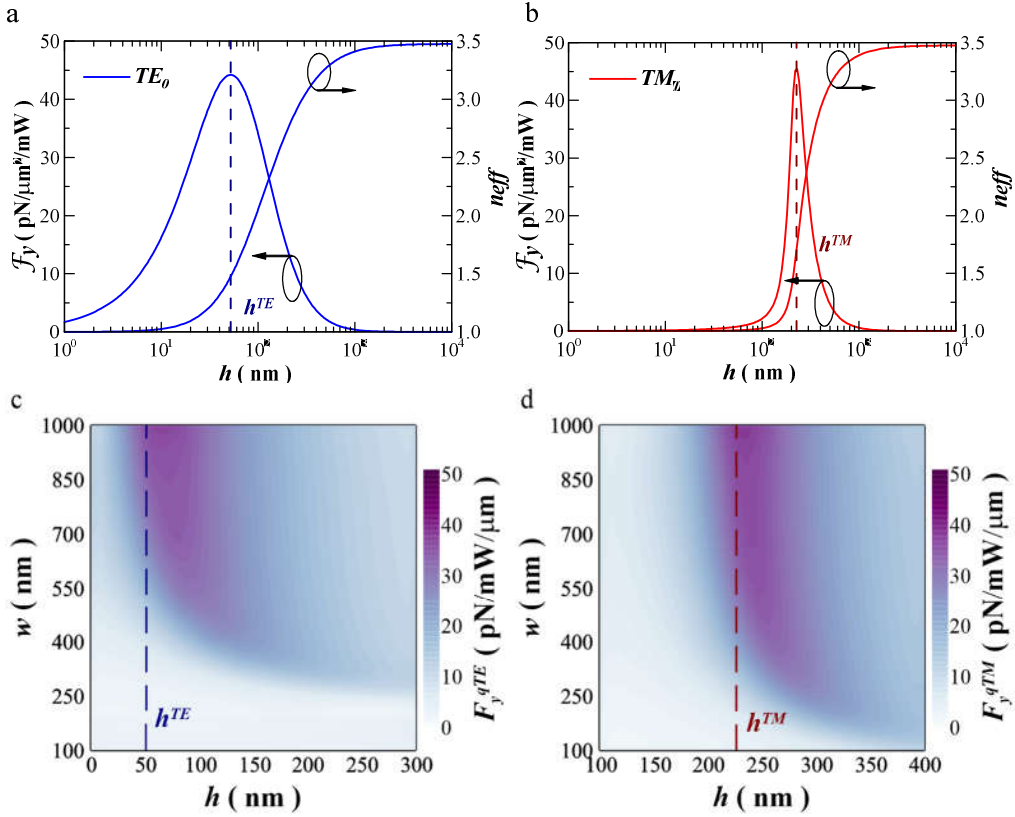

**Fig. 2. Optical forces distributions on the silicon waveguides in air and highlighted optimal height. (a)** and **(b)** Optical forces on the horizontal boundaries ($\mathcal{F}_y$) and their respective effective indexes for a silicon planar waveguide in air, for the fundamental TE and TM modes. **(c)** and **(d)** Intensity maps of y-component of optical forces ($F_y$) on the horizontal boundaries of the silicon rectangular waveguide in air, as a function of its cross-section dimension, for the quasi-TE and the quasi-TM fundamental modes at $\lambda_0 = 1550$ nm. The optimized heights are $h^{TE} = 52.04$ nm and $h^{TM} = 225.8$ nm.

On the other hand, Figs. 2(c)-(d) show the intensity maps of the y-component of the optical forces ($F_y$) applied over the horizontal boundaries of a silicon rectangular waveguide in air (Fig. 1(d)), as a function of its cross-section dimension, for both the quasi-TE (qTE) and the quasi-TM (qTM) fundamental modes at a wavelength of 1550 nm. These maps clearly indicate a strong convergence of the maximum optical forces to the same values of $h^{TE}$ and $h^{TM}$, also represented in Figs. 2(c)-(d). Although such tendencies could be expected in an extreme limit case, it is very surprising that they converge so quickly, taking into account the giant difference between their widths, where in the planar waveguide $w \to \infty$, whereas in the rectangular one, $w \cong 1$ μm. Another point we have noticed is that such a fast convergence has been achieved for an extreme high-index contrast case, as it is the the case for silicon in air, where the vector nature of the electromagnetic fields plays an important role; therefore, even better results are expected in the cases of lower index contrast. Additionally, the intensity maps also show that few tens nanometers off their optimized heights may represent a difference of almost 50-fold in th optical force magnitude, which can be tailored either to enhance or to suppress the Brillion gain.

To demonstrate the general applicability of the Eqs. (7) and (8) in optimizing the radiation pressure in realistic waveguides, we choose two very recent examples found in the literature. The first structure is composed by a silicon rib waveguide suspended in air, and the second one is a chalcogenide glass ($As_2S_3$) strip waveguide ($n_H = 2.4440$) buried in a silicon dioxide ($SiO_2$) ($n_L = 1.4440$), as illustrated schematically in Figs. 3(a) and (b), respectively. Both structures have successful used to experimentally demonstrate the Brillouin effect and its applications on integrated waveguide platforms [13,16,18,19,22,23], including the first demonstration of a silicon laser based on this effect [24]. Figures 3(c) and (d) present the total optical forces ($F_x + F_y$) for the silicon and the chalcogenide waveguides, respectively, for both polarizations at $\lambda_0 = 1550$ nm. The results show that the optimized heights computed by Eqs. (7) and (8) agree exactly with the points of maximum forces, which proves that in waveguides with high-aspect ratio of the type $w > h$, the y-component is the major responsible for the total optical force. Moreover, Fig. 3(c) shows that for silicon waveguide in air with a height near $h^{TM} = 225.8$ nm, the total optical force is approximately equal to 60 pN/μm/mW. This value corresponds to a total spatial-averaged radiation pressure ($\bar{p}_x + \bar{p}_y$) of 20.1x10³ N/m²/W in the fundamental qTM mode, whereas it is about 5.4x10³ N/m²/W (optical force of 15 pN/μm/mW) in the quasi-TE counterpart, which represents an enhancement of almost four times in optical pressure (force), just by changing the polarization. Similarly, Fig. 3(d) shows that by reducing the chalcogenide core height, from 850 nm to $h^{TE} = 119.1$ nm, increases the total radiation pressure from 294.9 to 3.6x10³ N/m²/W, keeping the same optical mode and polarization (qTE$_{00}$). Despite of the fact that Brillouin effects in nanowaguides have richer multiphysics, which involves the material photoelastic constants and mechanical modes dynamics, such results might represent significant improvements on their design. This is especially useful, if we take into account the fact that the Brillouin gain has a quadratic dependence on the optical force.

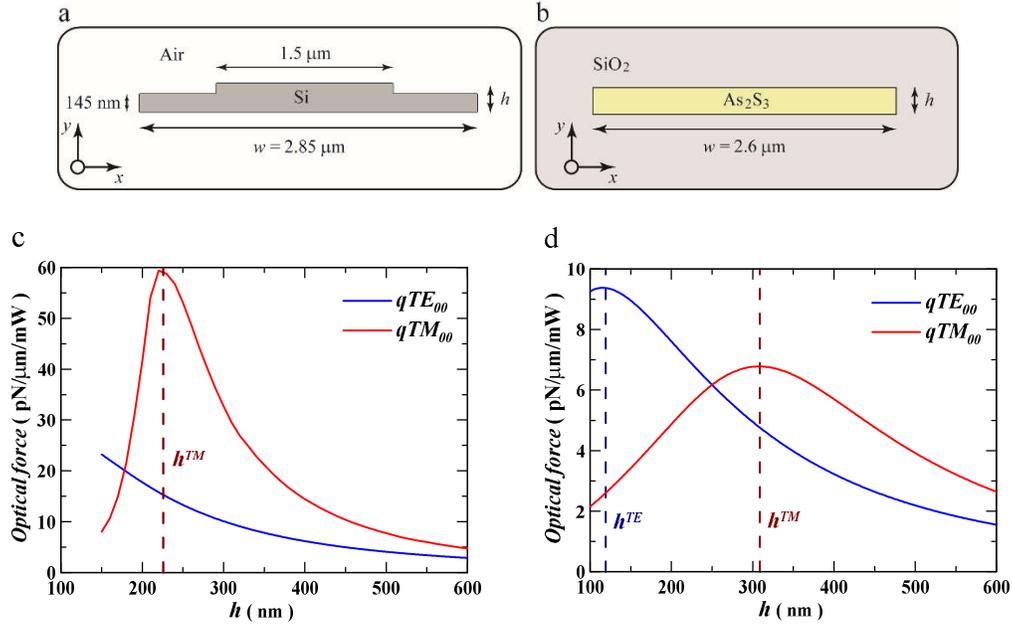

**Fig. 3. Schematic of waveguides and their respective total optical forces for both polarizations at a wavelength of 1550 nm.** (a) Silicon rib waveguide ($n_H = 3.4764$) suspended in air ($n_L = 1.0003$). (b) Chalcogenide strip waveguide ($n_H = 2.405$) buried in silicon dioxide cladding ($n_L = 1.4440$). (c) and (d) total optical forces on the silicon and the

chalcogenide waveguides, respectively. The optimized heights for the silicon waveguide are $h^{TE} = 52.04$ nm (not shown) and $h^{TM} = 225.8$ nm and for the chalcogenide are $h^{TE} = 119.1$ nm and $h^{TM} = 308.7$ nm.

The method developed here is based on a horizontal-oriented planar waveguide (*x*-direction invariance); therefore, the optimized height maximizes the *y*-component of the optical force ($F_y$). The numerical simulations show that for dielectric waveguides of the type $w > h$, when $w \geq 1$ μm, the derived expressions give the exact waveguide's height that maximizes the total optical force for any optical mode, polarization, and wavelength, covering most of the cases described in the literature. However, due to the waveguide symmetry and the orthogonality of the optical modes, orienting the planar waveguide in the vertical direction (*y*-direction invariance), our expressions may also be used to obtain its optimized width that now maximizes the *x*-component of the optical force ($F_x$). In the same way, it also gives exact results for waveguides of the type $w < h$, given that $h \geq 1$ μm. In fact, if we make the following changes: $w \to h$, $h \to w$, $F_y^{qTE} \to F_x^{qTM}$, $F_y^{qTM} \to F_x^{qTE}$ and $h^{TE,TM} \to w^{TM,TE}$, in the Figs. 2(c) and (d), they remain perfectly valid. We test the robustness of our findings by considering the fundamental qTE mode of a silicon rectangular waveguide with a fixed height of $h = 315$ nm surrounded by air at a wavelength of 1550 nm [10]. The specific width where the radiation pressure is maximum in the rectangular waveguide is $w = 273$ nm, whereas the optimized one is $w^{TM} = h^{TE} = 225.8$ nm; This small difference, in the tens-of-nanometers range, shows that our method still provides a quite good approximation, taking into account that, in this case, it is approximately a square waveguide ($w \sim h$).

Besides that, a fundamental question still remains: why do these very specific dimensions maximize the optical force? In order to explain such a behavior, we analyze Eq. (2), which shows that the radiation pressures reach their maximum values when the ratio $(n_g - n_{eff})/A_{wg}$ is maximum, for a given optical mode and polarization. Moreover, both the group and the effective indexes are intrinsically related to the total time-averaged electromagnetic (EM) energy per unit length of the waveguide $U_{Total}^{EM}$, which can be further separated into a transversal component $U_{xy}^{EM}$ and a longitudinal one $U_z^{EM}$, where $U_{Total}^{EM} = U_{xy}^{EM} + U_z^{EM}$. In that case, it is possible to show that the difference between $n_g$ and $n_{eff}$ relies on the fraction of energy in each component, more precisely $n_{eff} = n_g (U_{xy}^{EM} - U_z^{EM})/U_{Total}^{EM}$ [29-31]. In addition, the group index is related to the total EM energy by $n_g = c\, U_{Total}^{EM}/P$; By inserting these results in Eq. (2), we show that the radiation pressures reach their maximum values when the ratio $2U_z^{EM}/A_{wg}$ is maximum (see Supplementary Section B), revealing an implicit and strong dependence on the longitudinal component of the electromagnetic energy. Under this circumstance, the momentum flow in propagation direction, defined by $U_{xy}^{EM} - U_z^{EM}$, reaches its minimum value, resulting in the maximum radiation pressure value for a given waveguide area. It is also worth to notice that the longitudinal component of the EM energy ($U_z^{EM}$) depends solely on the longitudinal components of the EM fields ($E_z$, $H_z$) [30]. Consequently, as soon as the rectangular waveguide's width reaches few micrometers, it starts to behave exactly as a planar structure, maximizing the force in one transversal direction, what explains the fast convergence verified in Figs. 2(b) and (c). Another key point is that the TE polarization, in the planar waveguide approximation, has no electric field oriented in the propagation direction; thus, the longitudinal component of the EM energy is given exclusively by the magnetic energy - which reveals a hidden dependence of the optical force on the magnetic fields - a detail that has been overlooked so far.

In conclusion, we have used a planar dielectric waveguide and radiation pressure's theories to derive analytical expressions, capable of computing the exact waveguide's height (or width) that corresponds to the maximum value of the radiation-pressure induced optical force on the waveguide's boundaries, for any optical mode, polarization, and wavelength. We show that the origin of this point of maximum value of optical force (radiation pressure) is physically linked to the minimization of the momentum flow in the propagation direction. We believe that our findings are powerful and simple tools that can be directly applied to the optimization of Brillouin-based integrated waveguides and, therefore, this may have some impact on the designs of high-performance photonic-phononic technologies.

**Acknowledgments**
This work was supported in part by the Coordenação de Aperfeiçoamento de Pessoal de Nível Superior (CAPES) through a doctoral scholarship for J. R. Rodrigues and visiting professor sponsorship for V. R. Almeida, and in part by the Conselho Nacional de Desenvolvimento Científico e Tecnológico (CNPq) under Grant 310855/2016-0 and Grant 483116/2011-4.

# Supplementary Information

# Optimization of radiation pressure in dielectric nanowaveguides

**Janderson Rocha Rodrigues[1,2,*] and Vilson Rosa Almeida[1,2]**

[1]Instituto Tecnológico de Aeronáutica, 50 Praça Marechal Eduardo Gomes, Vila das Acacias, São José dos Campos - SP. 12228-900, Brazil
[2]Instituto de Estudos Avançados, 01 Trevo Cel Av José Alberto Albano do Amarante, Putim, São José dos Campos - SP. 12228-001, Brazil
[*]jrr@ita.br

**A. The derivative of the effective index with respect to the planar waveguide height.**

The even TE transcendental equation is given by:

$$tan\left(\kappa_H \frac{h}{2}\right) = \frac{\gamma_L}{\kappa_H} \tag{A1}$$

By performing the derivative on the right-hand side (RHS) of Eq. (A1) and using the trigonometric relation $sec^2\theta = 1 + tan^2\theta$, we have

$$RHS = \frac{d}{dw}\left[tan\left(\kappa_H \frac{h}{2}\right)\right] = sec^2\left(\kappa_H \frac{h}{2}\right)\frac{d}{dh}\left(\kappa_H \frac{h}{2}\right) =$$

$$= \left[1 + tan^2\left(\kappa_H \frac{h}{2}\right)\right]\frac{d}{dh}\left(\kappa_H \frac{h}{2}\right) = \frac{1}{2}\left(\frac{\kappa_H^2 + \gamma_L^2}{\kappa_H^2}\right)\left(-\frac{hk_0^2 n_{eff}}{\kappa_H}\frac{dn_{eff}}{dh} + \kappa_H\right) \tag{A2}$$

and on the LHS,

$$LHS = \frac{d}{dh}\left(\frac{\gamma_L}{\kappa_H}\right) = \frac{d\gamma_L}{dh}\frac{1}{\kappa_H} - \frac{d\kappa_H}{dh}\frac{\gamma_L}{\kappa_H^2}$$

$$= \frac{1}{2}\left(\frac{\kappa_H^2 + \gamma_L^2}{\kappa_H^2}\right)\frac{2k_0^2 n_{eff}}{\gamma_L \kappa_H}\frac{dn_{eff}}{dh} \tag{A3}$$

By putting together Eqs. (A2) and (A3) we obtain:

$$\frac{dn_{eff}}{dh} = \frac{\kappa_H^2}{n_{eff}k_0^2}\left(\frac{1}{h + 2/\gamma_L}\right) \tag{A4}$$

Substituting the definitions of $\kappa_H$ and $\gamma_L$ and manipulating it algebraically we arrive on Eq. (6). The exact solution described in Eq. (A4) is obtained for the TE antisymmetric transcendental equation,

$$tan\left(\kappa_H \frac{h}{2}\right) = -\frac{\kappa_H}{\gamma_L} \tag{A5}$$

Furthermore, the same procedures are applied for the symmetric TM transcendental equation:

$$tan\left(\kappa_H \frac{h}{2}\right) = \left(\frac{n_H}{n_L}\right)^2 \frac{\gamma_L}{\kappa_H} \tag{A6}$$

which results,

$$\frac{dn_{eff}}{dh} = \frac{\kappa_H^2}{n_{eff} k_0^2} \left\{ \frac{1}{w + \frac{2}{\gamma_L} \left[ \frac{\gamma_L^2 + \kappa_H^2}{\left(\frac{n_H}{n_L}\right)^2 \gamma_L^2 + \left(\frac{n_L}{n_H}\right)^2 \kappa_H^2} \right]} \right\} \tag{A7}$$

Again the exact same result of Eq. (A7) is obtained by deriving the following antisymmetric TM transcendental equation:

$$tan\left(\kappa_H \frac{h}{2}\right) = -\left(\frac{n_L}{n_H}\right)^2 \frac{\kappa_H}{\gamma_L} \tag{A8}$$

**B. The relation between the radiation pressure and longitudinal momentum flow**

The relation between the group velocity $v_g$, the phase velocity $v_p$, and the time-averaged EM energy per unit length is given by [29-31],

$$v_g = v_p \left\{ \frac{\int_{-\infty}^{+\infty} \int_{-\infty}^{+\infty} \varepsilon \left(|E_x|^2 + |E_y|^2 - |E_z|^2\right) + \mu \left(|H_x|^2 + |H_y|^2 - |H_z|^2\right) dxdy}{\int_{-\infty}^{+\infty} \int_{-\infty}^{+\infty} \varepsilon \left(|E_x|^2 + |E_y|^2 + |E_z|^2\right) + \mu \left(|H_x|^2 + |H_y|^2 + |H_z|^2\right) dxdy} \right\} \tag{B1}$$

However, the though the following relations: $v_g = c/n_g$ and $v_p = c/n_{eff}$, we can express Eq. (B1) as a function of the group index and the effective index,

$$n_{eff} = n_g \left\{ \frac{\int_{-\infty}^{+\infty} \int_{-\infty}^{+\infty} \varepsilon \left(|E_x|^2 + |E_y|^2 - |E_z|^2\right) + \mu \left(|H_x|^2 + |H_y|^2 - |H_z|^2\right) dxdy}{\int_{-\infty}^{+\infty} \int_{-\infty}^{+\infty} \varepsilon \left(|E_x|^2 + |E_y|^2 + |E_z|^2\right) + \mu \left(|H_x|^2 + |H_y|^2 + |H_z|^2\right) dxdy} \right\} \tag{B2}$$

where the transversal component of the EM energy is defined by

$$U_{xy}^{EM} = \int_{-\infty}^{+\infty}\int_{-\infty}^{+\infty} \varepsilon\left(|E_x|^2 + |E_y|^2\right) + \mu\left(|H_x|^2 + |H_y|^2\right) dxdy \tag{B3}$$

while the longitudinal one is

$$U_z^{EM} = \int_{-\infty}^{+\infty}\int_{-\infty}^{+\infty} \varepsilon|E_z|^2 + \mu|H_z|^2 dxdy \tag{B4}$$

Therefore, Eq. (B2) can be written by,

$$n_{eff} = n_g \left(\frac{U_{xy}^{EM} - U_z^{EM}}{U_{xy}^{EM} + U_z^{EM}}\right) \tag{B5}$$

Where the total EM energy per unit length is $U_{total}^{EM} = U_{xy}^{EM} + U_z^{EM}$. Therefore, by replacing Eq. (B5) in spatial-averaged radiation pressure for a rectangular waveguide we have,

$$\bar{p}_x + \bar{p}_y = \frac{P}{c}\frac{n_g}{A_{wg}}\left(\frac{2U_z^{EM}}{U_{total}^{EM}}\right) \tag{B6}$$

However, the total EM energy per unity length can be alternatively defined as,

$$U_{total}^{EM} = \frac{P}{v_g} = \frac{n_g P}{c} \tag{B7}$$

Substituting the Eq. (B7) into B6, we obtain:

$$\bar{p}_x + \bar{p}_y = \frac{2U_z^{EM}}{A_{wg}} \tag{B8}$$

Proving that the spatial-averaged of the radiation pressure, and therefore the induced optical force, can be expressed by the longitudinal component of the EM energy. On the other hand, the difference between the EM energy components $U_{xy}^{EM} - U_z^{EM}$ represents the EM momentum flow in the propagation direction (also given by the *zz*-component of the Maxwell stress tensor integrated over the waveguide cross section) [30]. Thus, the point of maximum on the radiation pressure ($U_z^{EM}$) represents a minimum for the momentum flow ($U_{xy}^{EM} - U_z^{EM}$) for a given waveguide area $A_{wg}$. Furthermore, as described in Eq. (B4), the longitudinal component of the EM energy depends on the longitudinal components of the fields ($E_z, H_z$).